\documentclass{article}
\usepackage{spconf,amsmath,graphicx,amssymb}

\graphicspath{{./figures/}}
\usepackage{url}
\usepackage{multirow}
\usepackage{balance}
\usepackage{subcaption}
\usepackage{cite}
\usepackage{booktabs}

\usepackage{color}

\title{UNSUPERVISED SPEAKER ADAPTATION USING ATTENTION-BASED\\SPEAKER MEMORY FOR END-TO-END ASR}
\name{Leda Sar{\i}$^{1,2}$, Niko Moritz$ ^1 $, Takaaki Hori$ ^1 $,  Jonathan Le Roux$ ^1 $\thanks{This work was done while L. Sar{\i} was an intern at MERL.}}
\address{$ ^1 $Mitsubishi Electric Research Laboratories (MERL), MA, USA\\
$ ^2 $Dept. of Electrical and Computer Engineering, University of Illinois at Urbana-Champaign, IL, USA}

\begin{document}
\ninept
\maketitle
\begin{abstract}
We propose an unsupervised speaker adaptation method inspired by the neural Turing machine for end-to-end (E2E) automatic speech recognition (ASR). The proposed model contains a memory block that holds speaker i-vectors extracted from the training data and reads relevant i-vectors from the memory through an attention mechanism. The resulting memory vector (M-vector) is concatenated to the acoustic features or to the hidden layer activations of an E2E neural network model. The E2E ASR system is based on the joint connectionist temporal classification and attention-based encoder-decoder architecture. M-vector and i-vector results are compared for inserting them at different layers of the encoder neural network using the WSJ and TED-LIUM2 ASR benchmarks. We show that M-vectors, which do not require an auxiliary speaker embedding extraction system at test time, achieve similar word error rates (WERs) compared to i-vectors for single speaker utterances and significantly lower WERs for utterances in which there are speaker changes.
\end{abstract}
\begin{keywords}
Unsupervised speaker adaptation, end-to-end speech recognition, neural Turing machine, speaker memory
\end{keywords}

\section{Introduction} \label{sec:intro}

Automatic speech recognition (ASR) models may not generalize well to mismatched speaker characteristics between the training and test data, which can lead to a significant degradation of the ASR accuracy.
In order to deal with this problem, various speaker adaptation methods have been proposed. These methods are generally crafted towards adapting Gaussian mixture model - hidden Markov model (GMM-HMM) systems \cite{gales1998maximum} and deep neural network (DNN) -HMM hybrids \cite{saon2013speaker,rath2013improved,abdel2013fast,swietojanski2014learning,cui2017embedding,sari2019learning}. Recently, end-to-end (E2E) systems that do not require explicit phonetic dictionaries have been used with increasing success for ASR. So far, however, only a few studies have dealt with speaker adaptation of E2E ASR models, which is the focus of this work. %

Prior studies on speaker adaptation of E2E systems include appending i-vectors to the acoustic features \cite{audhkhasi2017direct}, using speaker-transformed features obtained by feature space maximum likelihood linear regression (fMLLR) \cite{chorowski2014end}, using GMM-derived features \cite{tomashenko2018evaluation}, or using a speaker adversarial network \cite{meng2019adversarial}. Most of these methods apply adaptation only to the input features. However, at least for DNN-HMM systems, adaptation at the input layer has been shown not to be necessarily optimal \cite{kitza2018comparison}. %
It is thus worth investigating the effect of adaptation at different layers for E2E systems.

In this work, we propose to use the read mechanism of the neural Turing machine (NTM) \cite{graves2014neural} for speaker adaptation in E2E ASR. The read operation determines a weight distribution over all memory items, and each read memory vector, which we refer to as M-vector, is obtained as a weighted linear combination of i-vectors from the memory. The weights are computed using an attention mechanism and the M-vector is appended as speaker-adapting features. %
The main advantages of M-vector adaptation are that 
(a) it is unsupervised, in that it does not require adaptation data or additional label information such as speaker information at test time; %
(b) it is a frame-level adaptation approach, which enables faster adaptation to speaker changes and application to streaming ASR systems;
(c) any kind of embedding can be stored in the memory, even embeddings obtained from methods that may be impractical to be applied at test time in real-world systems; 
(d) the NTM interpretation of the adaptation opens the door to incorporating a write operation, i.e., learning speaker representations as we process the utterances, which is subject to further research.

The use of linear or convex combinations of existing speaker templates to represent new speakers or to adapt the model using a weighted combination of submodels has been studied in prior work.
However, none of these are applied to E2E systems. %
For example, \cite{kuhn1998eigenvoices} discusses the use of eigenvoices for speaker adaptation in GMM-HMM systems. In \cite{gales1998cluster}, the author uses a convex combination of the existing means for GMM mean adaptation. %
A linear combination of subnetworks is used in \cite{karanasou2017vectors} with a DNN-HMM system, which, however, requires adaptation data from the test speakers.
In \cite{delcroix2015context}, a weighted combination of the hidden layer activations is used, where the weights are determined by a context classifier. This method also differs from ours in that it does not use an external memory block interacting with the ASR network or attention-based reading. Recently, a weighted combination of speaker profiles has been used for speaker separation \cite{xiao2019single}. That work has however not been applied to speaker adaptation for ASR and it also does not provide an NTM interpretation of the model, which allows a more general adaptation scheme and provides opportunities for the introduction of a writable speaker memory that can adapt to and learn from new speakers.
Another study that makes use of the memory idea for ASR is \cite{zhang2016compact}, where the read vector is a convex combination of hidden layer states in the history rather than speaker representations and the memory is not applied for speaker adaptation. 

\begin{figure*}[t]
    \centering
    \includegraphics[scale=0.4250]{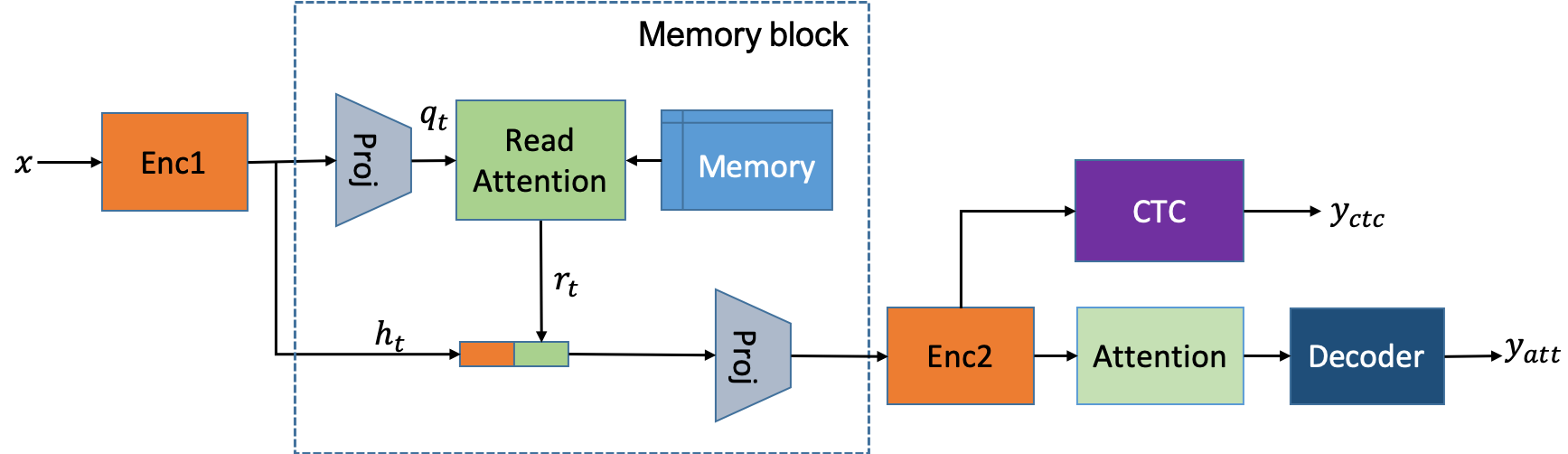}
    \caption{Block diagram of the E2E ASR system with memory-based adaptation network.}
   \label{fig:flow-memory}
   \vspace{-10pt}
\end{figure*}

We show that by using  M-vectors as speaker embeddings to append to the acoustic features, we achieve better performance than using the ``true'' i-vectors directly. In contrast with previous E2E speaker adaptation studies, we also investigate the effect of adapting intermediate layers of the encoder in an E2E ASR system, and show that it leads to lower word error rates (WERs) than adapting the input features.

The rest of the paper is organized as follows. Section \ref{sec:bgnd} summarizes the joint connectionist temporal classification (CTC) and attention-based encoder-decoder E2E ASR training, and the NTM. The proposed method is described in Section \ref{sec:method}. Experiments on WSJ and TED-LIUM2 are presented in Section \ref{sec:exps}. In Section \ref{sec:concl}, we summarize the paper and discuss possible future directions.

\vspace{-5pt}
\section{BACKGROUND} \label{sec:bgnd}

We first give a brief summary of E2E ASR systems that are trained using joint CTC and attention costs, and review the NTM concept with a focus on the read function. 

\subsection{Joint CTC and Attention E2E ASR}
E2E systems can be broadly classified into three groups, CTC-based systems~\cite{graves2014towards}, attention-based encoder-decoder models~\cite{chorowski2015attention}, and RNN transducer models \cite{graves2012sequence}, each aiming at achieving training with mismatched-length input-output pairs. Recently, CTC and encoder-decoder models have been combined using a multitask learning approach in order to mitigate their disadvantages~\cite{WatanabeHKHH17} and to enable streaming recognition of encoder-decoder architectures \cite{MoritzHR19,MoritzHR19c}.
Given an acoustic feature sequence $\mathbf{x}$ and the corresponding ground truth label sequence $\mathbf{y}^*$, the goal is to maximize the log-likelihood of the labels given the inputs in both approaches. However, they differ in their assumptions and the way the probabilities are computed. In CTC, the original label sequence is augmented with blank symbols to generate $\mathbf{y}'$ and the softmax output $s_t$ of the DNN at time $t$ along with forward ($\alpha_t$) and backward ($\beta_t$) probabilities of partial hypotheses $\mathbf{y}_{1:u}'$, are used to compute the CTC loss
\begin{equation}
\mathcal{L}_{\text{ctc}} = - \log P_{\text{ctc}}(\mathbf{y}^* | \mathbf{x}) =  - \log \sum_{u=1}^{|\mathbf{y}'|} \frac{\alpha_t(u)\beta_t(u)}{s_t(\mathbf{y}_u')}, \label{eq:ctc}
\end{equation}
where $|\mathbf{y}'|$ denotes the length of the augmented label sequence $\mathbf{y}'$.
In the attention-based approach, the conditional probability $P_\text{att}(y_u^* | \mathbf{x}, y^*_{1:u-1})$ is the output of the decoder model and the loss function $\mathcal{L}_\text{att}$ is derived as follows
\begin{equation}
    \mathcal{L}_{\text{att}} = - \log P_{\text{att}}(\mathbf{y}^* | \mathbf{x}) = - \log \sum_u  P_{\text{att}}(y_u^* | \mathbf{x}, y^*_{1:u-1}). \label{eq:att} 
\end{equation}
In the joint training framework, these two losses are combined using a weight parameter $\lambda$:
\begin{align}
    \mathcal{L}_{\text{joint}} &= \lambda \mathcal{L}_{\text{ctc}} + (1-\lambda) \mathcal{L}_{\text{att}} . \label{eq:mtl}
\end{align}

\subsection{Neural Turing Machine}
The NTM is a differentiable computer that has a memory component, read and write attention heads to interact with the memory, and a controller to determine the actions of these heads \cite{graves2014neural}. Let $M$ be the memory of the NTM, represented by a $ D \times N $ matrix, where $N$ is the number of items in the memory and $D$ the size of the memory vectors. Given a query vector $ q_t \in \mathbb{R}^D $ at time $ t $, the memory reading step consists in computing a weighted sum $r_t$ over the memory items $M_n$, with weights $w_t(n)$ determined by the cosine similarity between the query and the items modulated by a scalar $\gamma_t$ as follows:  %
\begin{align}
K(q_t, M_n) &= \frac{q_t^\top M_n}{||q_t||\,||M_n||}, \label{eq:dot}\\
w_t(n) &= \frac{e^{\gamma_t K(q_t, M_n)}}{\sum_{n'=1}^{N}e^{\gamma_t K(q_t, M_{n'})}}, \label{eq:wt}\\
r_t &= \sum_{n=1}^{N} w_t(n) M_n. \label{eq:rt}
\end{align}
In this work, we only utilize the read head, hence we shall skip the description of the writing mechanism. The interested readers can refer to~\cite{graves2014neural} for more details.

\section{MEMORY BASED ADAPTATION FOR E2E ASR}
\label{sec:method}

The structure of our memory-based adaptation network is shown in Fig.~\ref{fig:flow-memory}. In this architecture, the encoder is split into two parts (Enc1 and Enc2), where each part can include zero or more layers. The query vector $q_t$ is computed by applying a projection layer on the output of the first encoder. The query vector is used to read from the speaker memory and to determine the speaker memory vector (M-vector) $r_t$ using Eq.~\eqref{eq:rt}. As the query vector is time-dependent, %
the computed M-vector changes at each frame as well, allowing the flexibility to change the adaptation parameters within an utterance. The M-vector is appended to the Enc1 output $h_t$ and projected to a lower dimensional vector using another projection layer. This output is fed into the second part of the encoder Enc2, whose output is fed into both the CTC module and the attention module followed by a decoder module, which is known as a joint CTC-attention based E2E ASR system.

In this work, speaker i-vectors from the training data are used as the memory elements. Instead of the cosine similarity of Eq.~(\ref{eq:dot}) as used in the NTM, the scaled dot-product based attention mechanism is applied to read from the memory instead, where the scale is the square root of the vectors dimension \cite{vaswani2017attention}. We also set $\gamma_t = 1$ in Eq. \ref{eq:wt}.  %
In these equations, the output dimension of the projection that generates the query must match the dimensionality $D$ of the memory vectors. However, the proposed approach is not limited to this case, as other attention methods without restriction on the dimension of $q_t$ can be used. 

In the training phase, speaker i-vectors of each training speaker are extracted and arranged in a matrix to form the speaker memory. The read mechanism of the memory block, the encoder, and the decoder are trained using the joint CTC-attention objective~\cite{WatanabeHKHH17}. During test time, the memory block remains unaltered and only contains speaker i-vectors extracted from the training data. Since no additional speaker information is extracted on the test data, the proposed method achieves unsupervised adaptation. %

\section{EXPERIMENTS}
\label{sec:exps}
We investigate the performance of the proposed approach on two datasets, the Wall Street Journal (WSJ) corpus of read English newspapers~\cite{wsj0,wsj1}, and the TED-LIUM2 corpus of TED conference talks~\cite{tedlium2}. Training, development, and test set sizes in hours are 81.3, 1.1, and 0.7 for WSJ, and 211.1, 1.6, and 2.6 for TED-LIUM2, respectively. We use TED-LIUM2 as a second dataset to investigate how the performance evolves when using a larger training set with a higher number of speakers. We use the ESPnet toolkit~\cite{watanabe2018espnet} to implement and investigate our proposed methods. I-vectors are extracted using the Kaldi toolkit~\cite{povey2011kaldi}. For the baseline systems without adaptation, we use the default ESPnet recipes. 
We compare the performance of the proposed M-vector approach with similar systems in which either the oracle speaker i-vector or utterance-level i-vectors are repeated and appended at each frame.
Training of each ASR system setup is conducted four times using a different random seed for initializing neural network weights, and results of the model with the best development set performance are reported.

We use 80-dimensional mel filterbank energies as input features along with pitch features, namely probability of voicing, log-pitch, and delta-log-pitch. The E2E model is character-based, and output units consist of letters of the English alphabet and symbols for blank, unknown, quote, space, end-of-sentence, and punctuation marks.

In the i-vector extraction, the trained universal background model (UBM) has 1024 components and the i-vector dimension is 100. The speaker memory is of size $D \times N$ with $D$ being the i-vector dimension and $N$ the number of speakers in the training dataset.

\vspace{-5pt}
\subsection{WSJ Experiments}
\label{sec:wsj}
The baseline E2E ASR model uses a stack of 6 encoder layers each consisting of a BLSTM followed by a projection layer with 320 units in each direction. The decoder has a single LSTM layer with 300 units. Location-based attention is used between the encoder and the decoder. In joint CTC-attention training, the weight for the CTC loss in Eq.~\eqref{eq:mtl} is set to $\lambda = 0.2$. At recognition time, a recurrent neural network-based word language model (RNN-LM), which consists of a single LSTM layer with 1000 units, is used via shallow fusion \cite{hori2018end}. Decoding is performed with a beam size of 30. The memory size $N$ for WSJ is 283, which is equal to the number of speakers in the training set. The development and evaluation sets include 10 and 8 speakers, respectively, who are distinct from the training speakers.

Table~\ref{tab:wsj-baseline} shows results when appending i-vectors or M-vectors at a given encoder neural network layer $l$, with $0\leq l \leq 6$.
When $l=0$, the vectors are appended to the input features, otherwise to the output of the $l$-th encoder layer (referring to the notations of Fig.~\ref{fig:flow-memory}, Enc1 then consists of the first $l$ layers, and Enc2 of the remaining $6-l$ layers). %
The i-vector system uses oracle speaker i-vectors. ASR results without speaker adaptation are also shown for comparison.

\begin{table}[t]
    \centering
    \caption{WERs [\%] on the WSJ task for the i-vector and M-vector based systems that adapt the output of the $l$-th layer of the encoder ($l=0$ denotes input features).}
    \label{tab:wsj-baseline}
    \vspace{-0.1cm}
    \begin{tabular}{ccccc}
    \toprule
    & \multicolumn{2}{c}{i-vector} &  \multicolumn{2}{c}{M-vector}\\ 
    \cmidrule(lr){2-3} \cmidrule(lr){4-5} 
          Layer & dev93 & eval92& dev93 & eval92 \\ \midrule
         $l=0$ & 6.8& 4.4 & 7.3& 4.2 \\
         $l=1$ & 6.7& 4.4 & 6.6& 4.2\\
         $l=2$ &  6.5 &  4.7 & {\bf 6.5}& {\bf 4.2} \\
         $l=3$ & {\bf6.4}& {\bf4.7} & 6.8& 4.6 \\
         $l=4$ & 7.0& 4.6 & 7.3& 4.3\\
         $l=5$ & 6.6& 4.3 & 7.2& 4.6\\
         $l=6$ & 7.2& 4.7 & 7.9& 4.4 \\ \midrule 
         No adaptation & 8.9 & 5.8 & 8.9 & 5.8 \\
        \bottomrule
    \end{tabular}
\end{table}

We first see that both the i-vector and M-vector approaches reduce the WERs significantly, implying that speaker information can be effectively used to adjust the model within our E2E ASR system.
Furthermore, the proposed unsupervised adaptation approach using M-vectors obtains comparable performance with the supervised i-vector baseline on the development set, and a relative improvement of 10.6\% on the evaluation set. This demonstrates that the proposed memory-based approach is applicable to unseen test speakers. %
Note that, despite the apparent advantage of using the true speaker i-vector computed on the test set, the i-vector baseline sees the same information repeated for each frame in the utterance, whereas the proposed approach has the opportunity to take advantage of frame-level adaptation, thus achieving a lower WER. 

If we compare performance depending on the layer at which adaptation is performed, for i-vectors, layer 3 gives the lowest WERs, and for M-vectors, adaptation after the second layer gives the best results while performance tends to degrade as $l$ gets larger.  

We shall stress here the fact that the i-vector baseline requires speaker labels for the test set, and thus is a form of speaker-aware adaptation, contrary to our proposed unsupervised adaptation method. In order to eliminate this obvious advantage of the speaker i-vector approach, we also tested the system with utterance-based test set i-vectors, which are shown in the middle two columns of Table~\ref{tab:wsj-utt-dec}. In all cases, we observed an increase in WER on the test set mainly because of the mismatch between training and testing conditions (speaker vs.\ utterance). To prevent this mismatch, we also tested the condition where utterance i-vectors are used for both training and testing, which is shown in the last two columns of Table~\ref{tab:wsj-utt-dec}. If we compare the utterance-level i-vector results with the M-vector approach shown in Table~\ref{tab:wsj-baseline}, we see that M-vectors achieve 10.6\% (4.7$\rightarrow$4.2) relatively lower WER on the evaluation set while having a similar development set performance. %

\begin{table}[t]
    \centering
    \caption{WERs [\%] on WSJ for i-vector based systems when training/testing with speaker (Spk) or utterance-level (Utt) i-vectors.} 
    \label{tab:wsj-utt-dec}
    \vspace{-0.1cm}
    \begin{tabular}{ccccccc}
    \toprule
    Train & \multicolumn{2}{c}{Spk i-vec} & \multicolumn{2}{c}{Spk i-vec} & \multicolumn{2}{c}{Utt i-vec} \\ \midrule
     Test & \multicolumn{2}{c}{Spk i-vec} & \multicolumn{2}{c}{Utt i-vec} & \multicolumn{2}{c}{Utt i-vec}  \\ \midrule
         & dev & test & dev & test & dev & test \\ \midrule
         $l=1$ & 6.7 & 4.4 & 7.2 & 4.9  & 6.8 & 4.4 \\
         $l=2$ & 6.5 & 4.7 & 6.3 & 4.8  & 6.6 & 4.3\\ 
         $l=3$ & 6.4 & 4.7 & 6.4 & 4.9  & \textbf{6.4} & \textbf{4.7} \\
         \bottomrule
    \end{tabular}
    \vspace{-10pt}
\end{table}

In order to show the advantage of the frame-level adaptation strategy of the M-vector approach, we also tested the models trained with single-speaker utterances on utterances with speaker change. The speaker change condition is synthetically generated by sampling pairs of utterances from different speakers and concatenating them while ensuring that each test utterance is used exactly once. In addition, silence parts between concatenated utterances are removed.

\begin{table}[t]
\centering
\caption{WERs [\%] on single speaker utterances (dev93, eval92) and simulated speaker change utterances (dev93$^*$, eval92$^*$) from WSJ.}
\label{tab:wsj-spk-change}
\vspace{-0.1pt}
    \begin{tabular}{ccccc}
    \toprule
    & \multicolumn{2}{c}{single speaker} & \multicolumn{2}{c}{speaker change} \\
    \cmidrule(lr){2-3} \cmidrule(lr){4-5}
         & ~dev93~ & ~eval92~ & ~~dev93$^*$~~ & ~~eval92$^*$~~ \\ \midrule
        i-vector & {\bf 6.4} & 4.7 & 10.4 & 7.8 \\ 
        M-vector & 6.5 & {\bf 4.2} & {\bf 7.6} & {\bf 4.9} \\ \bottomrule
    \end{tabular}
\end{table}

For the i-vector setup, we extracted i-vectors using the concatenated utterances and the performances of the best i-vector ($l=3$) and M-vector ($l=2$) setups are compared for these new test sets as shown in Table~\ref{tab:wsj-spk-change}.
As we can see, WER for the i-vector system increases by up to 4\% absolute, whereas WER for the M-vector system increases only by 1\%, showing that the M-vector based system is more robust to speaker change condition. %

\subsection{TED-LIUM2 Experiments}
The baseline E2E model and the RNN-LM structures for the TED-LIUM2 experiments are similar to the WSJ setup except that the mixing weight for the CTC loss is set to $\lambda = 0.5$. The memory size $N$ for the TED-LIUM2 experiments is 1267. While the default Kaldi setup considers different recordings of the same speaker as different speakers, we here mapped all those recordings to the same speaker. %
The development and test sets include 8 and 11 speakers, respectively. %

Table~\ref{tab:ted-baseline} shows results of the speaker i-vector and M-vector based systems, as well as of the unadapted model. %
Based on the observation that the adaptation is not effective when $l$ gets larger, we performed TED-LIUM2 experiments with adaptation only up to and including the third layer. As shown in Table~\ref{tab:ted-baseline}, appending the i-vectors to the first layer ($l=1$) leads to lower development set WER than using layers $l=2$ or $l=3$. %
\begin{table}[t]
    \centering
    \caption{WERs [\%] on the TED-LIUM2 task for the i-vector and M-vector based systems that adapt the output of the $l$-th layer of the encoder ($l=0$ denotes input features).}
    \label{tab:ted-baseline}
    \vspace{-0.1pt}
    \begin{tabular}{ccccc}
    \toprule
    & \multicolumn{2}{c}{i-vector} &  \multicolumn{2}{c}{M-vector}\\ 
    \cmidrule(lr){2-3} \cmidrule(lr){4-5}
         Layer & dev & test & dev & test \\  
         \midrule
         $l=0$ & 12.9 & 12.3 & 12.3 & 11.7\\ 
         $l=1$ & {\bf 11.7} & {\bf 11.2} & 12.1 & 11.6\\ %
         $l=2$ & 12.0 & 11.6 & 12.1 & 11.7 \\
         $l=3$ & 11.8 & 11.4 & {\bf 11.8} & {\bf 11.0}  \\ \midrule
         No adaptation & 18.6  & 16.7 & 18.6  & 16.7\\
        \bottomrule
    \end{tabular}  
\end{table}

The best WERs for the proposed M-vector approach are obtained for $l=3$. In this setup, M-vectors perform about equally well compared to the best i-vector results of Table~\ref{tab:ted-baseline}. Our approach still provides the advantage of the unsupervised online adaptation as compared to the i-vector baseline as we do not need to compute the test set i-vectors and simply use the training memory for adaptation. Moreover, the speaker i-vector system here is allowed to see all utterances of the same speaker, whereas our method is agnostic to this.

In Table~\ref{tab:ted-utt-dec}, we compare speaker and utterance-level i-vectors for the TED-LIUM2 dataset. %
Using utterance-level i-vectors increases the test set WER by 0.7\% absolute compared to using speaker-level i-vectors.
Compared to M-vector results shown in Table~\ref{tab:ted-baseline}, we see that the M-vector approach applied to $l=3$ achieves a similar development set performance%
, and a 7.6\% (11.9$\rightarrow$11.0) relatively lower WER for the test data set. If utterance i-vectors are used during training as well, i.e., without training mismatch, the i-vector test set WER increases by 0.6\% absolute as compared to the speaker i-vector model.
If we compare this to the best M-vector result from Table \ref{tab:ted-baseline}, we observe 6.8\% (11.8$\rightarrow$11.0) relatively lower WER than the utterance i-vector based system. 

\begin{table}[t]
    \centering
    \caption{WERs [\%] on TED-LIUM2 for i-vector based systems when training/testing with speaker (Spk) or utterance-level (Utt) i-vectors.} %
    \label{tab:ted-utt-dec}
    \vspace{-2pt}
    \begin{tabular}{ccccccc}
    \toprule
    Train & \multicolumn{2}{c}{Spk i-vec} & \multicolumn{2}{c}{Spk i-vec} & \multicolumn{2}{c}{Utt i-vec} \\ \midrule
    Test & \multicolumn{2}{c}{Spk i-vec} & \multicolumn{2}{c}{Utt i-vec} & \multicolumn{2}{c}{Utt i-vec} \\ \midrule
         & dev & test & dev & test & dev & test\\ \midrule
         $l=1$ & \textbf{11.7} & \textbf{11.2} &\textbf{ 11.7 }& \textbf{11.9}  & \textbf{11.7}& \textbf{11.8} \\
         $l=2$ & 12.0& 11.6 & 12.2 & 12.3 & 11.9& 11.8\\ 
         $l=3$ & 11.8 & 11.4 & 12.1 & 12.0 & 11.8& 11.9 \\
         \bottomrule
    \end{tabular}
\end{table}

We also compared the performance of M-vectors and i-vectors with speaker changes on the TED-LIUM2 data set by synthetically concatenating utterances from different speakers as described in Section \ref{sec:wsj} and decoding using the best i-vector and M-vector models. The resulting WERs are given in Table~\ref{tab:ted-spk-change}. We can see that M-vectors are more robust to the speaker change scenario than the i-vectors. On the test set, for example, we achieve 4\% absolute lower WER than the i-vector system, which demonstrates the advantage of frame-level adaptation.   

\begin{table}[t]
\centering
\caption{WERs [\%] on single speaker utterances (dev, test) and simulated speaker change utterances (dev$^*$, test$^*$) from TED-LIUM2.}
\label{tab:ted-spk-change}
\vspace{-2pt}
    \begin{tabular}{ccccc}
   \toprule
    & \multicolumn{2}{c}{single speaker} & \multicolumn{2}{c}{speaker change} \\
    \cmidrule(lr){2-3} \cmidrule(lr){4-5} 
         & ~~dev~~ & ~~test~~ & ~~dev$^*$~~ & ~~test$^*$~~ \\ \midrule
        i-vector & {\bf 11.7} & 11.2 & 16.1 & 15.9 \\ 
        M-vector & 11.8 & {\bf 11.0} & {\bf 14.1} & {\bf 11.9} \\ \bottomrule
    \end{tabular}
\end{table}

\section{CONCLUSIONS AND FUTURE WORK}
\label{sec:concl}
In this work, we presented an NTM-inspired speaker adaptation method for E2E ASR systems. We proposed to use a speaker i-vector memory that interacts with the encoder using an attention mechanism. The attention mechanism determines the mixing weights that are used to combine memory i-vectors. The read memory vector, or M-vector, is appended to the acoustic features or to the output of an intermediate encoder layer and then projected before being fed into the next layer. %
We showed that, compared with speaker i-vectors and utterance i-vectors, the proposed M-vector system achieved similar or improved WERs on the WSJ and TED-LIUM2 datasets despite the fact that the M-vector approach is unsupervised and does not require i-vector computation at test time.
We also showed that M-vectors are significantly more robust to the case where test utterances contain a speaker change.

Future directions include investigating the effect of different types of speaker embeddings such as x-vectors~\cite{snyder2018x} or d-vectors~\cite{variani2014deep}.
Going further in the application of the NTM concept, incorporating a writing capability for the speaker memory can be explored. This way, the network can gradually learn to store previously seen speakers. %

\vfill\pagebreak
\balance

\bibliographystyle{IEEEtran}
\bibliography{refs}

\end{document}